\documentclass[preprint2,twoside]{hwo}

\usepackage{graphicx}
\usepackage[normalem]{ulem}
\usepackage{enumerate}
\usepackage{etaremune}
\usepackage{enumitem}
\usepackage{makecell}

\graphicspath{{./}{figures/}}

\newcommand{\smallspace}{-0.05cm}

\input{hwo.h}

\begin{document}

\title{\textbf{\LARGE Detecting Surface Liquid Water on Exoplanets}}

\author {\textbf{\large Nicolas B. Cowan$^{1,2}$}}
\affil{$^1$\small\it Department of Physics, McGill University, 3600 rue University, Montréal QC H3A 2T8, Canada}
\affil{$^2$\small\it Department of Earth \& Planetary Sciences, McGill University, 3450 rue University, Montréal, QC H3A 0E8, Canada} 

\author {\textbf{\large Jacob Lustig-Yaeger$^{3,4}$}}
\affil{$^3$\small\it JHU Applied Physics Laboratory, 11100 Johns Hopkins Rd, Laurel, MD 20723, USA}
\affil{$^4$\small\it NASA NASA Nexus for Exoplanet System Science, Virtual Planetary Laboratory, Box 351580, University of Washington, Seattle, WA 98195, USA}

\author {\textbf{\large Renyu Hu$^{5,6}$}}
\affil{$^5$\small\it Jet Propulsion Laboratory, California Institute of Technology, Pasadena, CA 91011, USA} 
\affil{$^6$\small\it Division of Geological and Planetary Sciences, California Institute of Technology, Pasadena, CA 91125, USA} 

\author {\textbf{\large L.~C. Mayorga$^{3}$}}

\author {\textbf{\large Tyler D. Robinson$^{4,7,8}$}}
\affil{$^7$\small\it Lunar \& Planetary Laboratory, University of Arizona, Tucson, AZ 85721, USA}

\affil{$^8$\small\it Habitability, Atmospheres, and Biosignatures Laboratory, University of Arizona, Tucson, AZ 85721, USA}

\author {\textbf{\large on behalf of Characterizing Exoplanets sub working group}}

\author{\footnotesize{\bf Endorsed by:}
Eleonora Alei (NASA Goddard Space Flight Center), 
Natalie Allen (Johns Hopkins University), 
David Arnot (The Open University), 
Reza Ashtari (JHU-APL), 
Katherine Bennett (Johns Hopkins University), 
Jayne Birkby (University of Oxford), 
Abby Boehm (Cornell University), 
Kara Brugman (Arizona State University), 
Aarynn Carter (STScI), 
Katy L. Chubb (University of Bristol), 
Luca Fossati (Space Research Institute, Austrian Academy of Sciences), 
Theodora Karalidi (University of Central Florida), 
Finnegan Keller (Arizona State University), 
James Kirk (Imperial College London), 
Joshua Krissansen-Totton (University of Washington), 
Alvaro Labiano (Telespazio UK), 
Eunjeong Lee (EisKosmos (CROASAEN), Inc.), 
Briley Lewis (University of California Santa Barbara), 
Evelyn  Macdonald (University of Vienna), 
Stanimir Metchev (University of Western Ontario), 
Blair Russell (Chapman University), 
Arnaud Salvador (German Aerospace Center), 
Gaetano Scandariato (INAF), 
Everett Schlawin (University of Arizona), 
Jessica Spake (Carnegie Observatories), 
Sarah Steiger (Space Telescope Science Institute), 
Tomas Stolker (Leiden University), 
Johanna Teske (Carnegie Earth and Planets Lab), 
Thaddeus Komacek (University of Oxford), 
Martin Turbet (LMD, LAB, IPSL, CNRSS), 
Daniel Valentine (University of Bristol), 
Hannah R. Wakeford (University of Bristol), 
Austin Ware (Arizona State University), 
Thomas Wilson (University of Warwick), 
Nicholas Wogan (NASA Ames Research Center) 
}



\begin{abstract}
Planets with large bodies of water on their surface will have more temperate and stable climates, and such planets are the ideal places for life-as-we-know-it to arise and evolve. A key science case for the Habitable Worlds Observatory (HWO) is to determine which planets host surface liquid water. Aside from its implications for planetary climate and astrobiology, detecting surface water on terrestrial exoplanets would place important constraints on our theories of planet formation and volatile delivery. Rotational variability in the reflectance of an exoplanet may reveal surface features rotating in and out of view, including oceans. Orbital changes in reflectance and polarization, meanwhile, are sensitive to the scattering phase function of the planetary surface, including specular reflection from large bodies of water. Although these techniques are applicable to all temperate terrestrial exoplanets, we focus in this document on the directly-imaged planets that are more likely to drive the HWO coronagraph design. Identification of water oceans relies on detecting a liquid, and using other lines of evidence to narrow that liquid down to being water. Liquids have smoother surfaces than most solids, and hence exhibit specular reflection instead of diffuse reflection. In practice, this makes lakes and oceans look dark from most illumination angles, but mirror-like at glancing angles. HWO is uniquely capable of identifying surface liquid oceans via their optical properties. Given that discovering an ocean on an exoplanet would confirm its status as a habitable world, this science case is literally the \emph{raison d’être} of the Habitable Worlds Observatory.  
(This article is an adaptation of a science case document developed for HWO's Solar Systems in Context, Characterizing Exoplanets Steering Committee.)
\end{abstract}


\section{Science Goal}

The fundamental question we aim to address in this HWO science case is: \\ 

\noindent \textbf{\textit{How common are oceans on habitable zone rocky planets?}} \\ 

\noindent Planets with large bodies of water at their surface will have more temperate and stable climates, and such planets are the ideal places for life-as-we-know-it to arise and evolve \citep{10.3389/fspas.2025.1544426}. A key science case for the Habitable Worlds Observatory is therefore to determine which planets host surface liquid water. Aside from its implications for planetary climate and astrobiology, detecting surface water on terrestrial exoplanets would place important constraints on our theories of planet formation and volatile delivery \citep{2020plas.book..325M}. 

\section{Science Objective}

\noindent \textbf{\textit{Determine the presence of an ocean on an exoplanet by spatially mapping its surface or detecting ocean glint.}} \\ 

Rotational variability in the reflectance of an exoplanet may reveal surface features rotating in and out of view, including oceans. Orbital changes in reflectance and polarization, meanwhile, are sensitive to the scattering phase function of the planetary surface, including specular reflection from large bodies of water. Although these techniques are in principle applicable to all temperate terrestrial exoplanets, including those on short orbits around red dwarfs, we focus in this document on the directly-imaged planets that are more likely to drive the HWO coronagraph design.

\begin{figure*}[bth!]
    \centering
    \includegraphics[width=0.95\textwidth]{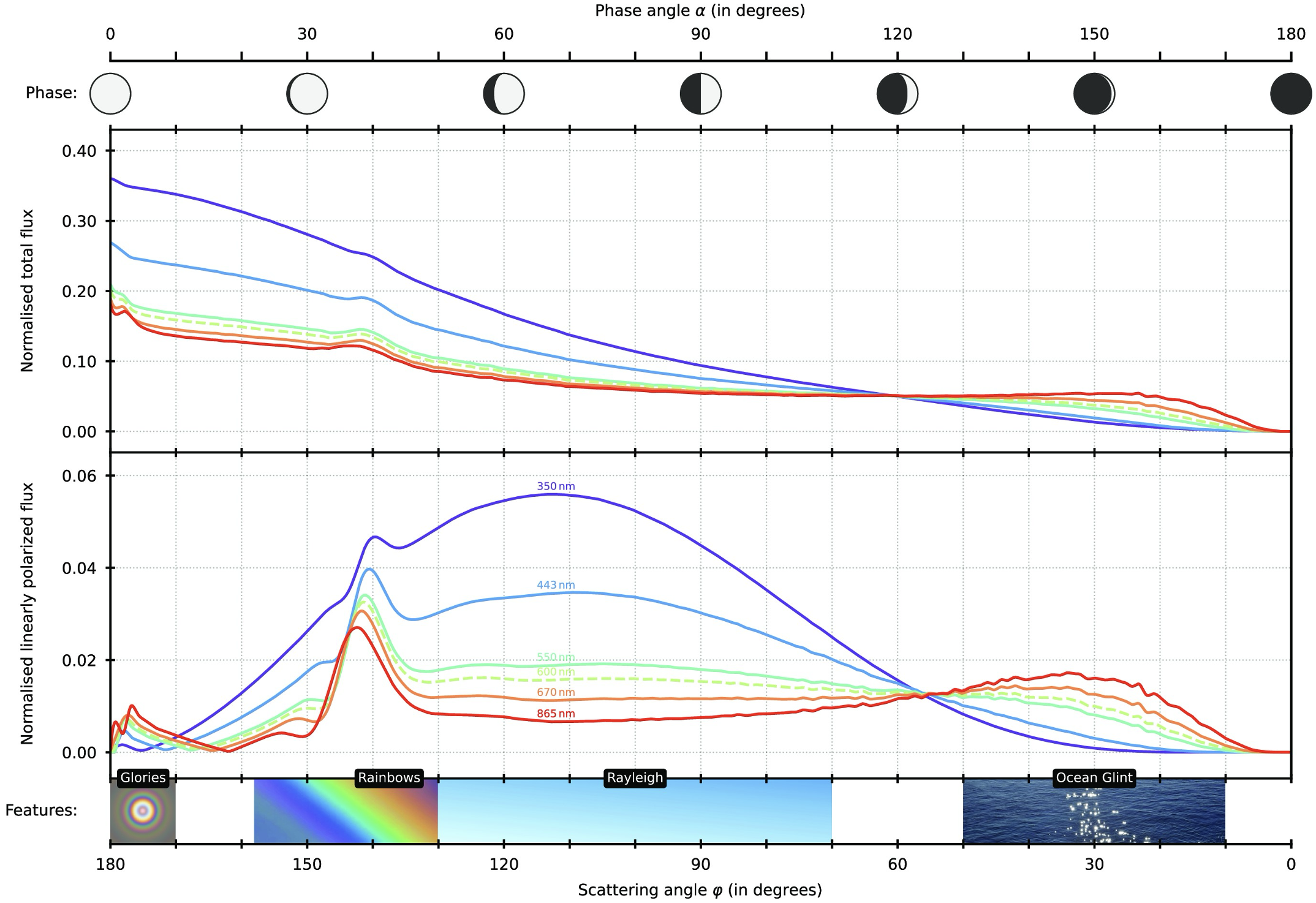}
    \caption{Planetary reflectivity (\emph{top}; total flux) and polarization (\emph{bottom}; polarized flux) as a function of scattering angle (or phase angle) and wavelength (colors). Glories, rainbows, Rayleigh scattering, and ocean glint generate strong polarization signatures and subtle reflectance signatures. \citep[Figure from][]{Vaughan2023}.
}
    \label{fig:fig1}
\end{figure*}

Identification of water oceans relies on detecting a liquid, and using other lines of evidence to narrow that liquid down to being water. Liquids have smoother surfaces than most solids, and hence exhibit specular reflection instead of diffuse reflection. In practice, this makes lakes and oceans look dark from most illumination angles, but mirror-like at glancing angles (e.g., methane lakes on Titan: \citealp{Stofan2007, Barnes2014}). Surface oceans can therefore be identified as the dark regions when the surface of a planet is rotationally mapped near quadrature or gibbous phases \citep{Cowan2009, Cowan2011} and then appear reflective —and linearly polarizing— at crescent phases \citep{Robinson2010, Lustig-Yaeger2018, Vaughan2023}. To summarize, liquid water on the surface of an exoplanet can be identified in three complementary ways: dark regions at gibbous phases, glint and linear polarization at crescent phases.

Clouds present a challenge in all cases: one can only map the surface of cloud-free regions of a planet, and time varying clouds greatly complicate the mapping exercise.  Forward scattering of clouds at crescent phases, meanwhile, can masquerade as ocean glint, although polarization or spectroscopy can help distinguish cloud scattering from glint effects \citep{Trees2019, Ryan2022}. In other words, clouds act as both a false negative, by masking surface water, and as a false positive, by mimicking specular reflection. Unfortunately, clouds almost certainly accompany oceans of liquid water, so this is a confounding factor that we will have to plan for and mitigate. Water makes up 71\% the surface of the Earth, while 50--70\% of the planet is covered in clouds at any given moment. 

\subsection{Ocean Glint}

\begin{figure*}[th!]
    \centering
    \includegraphics[width=0.95\textwidth]{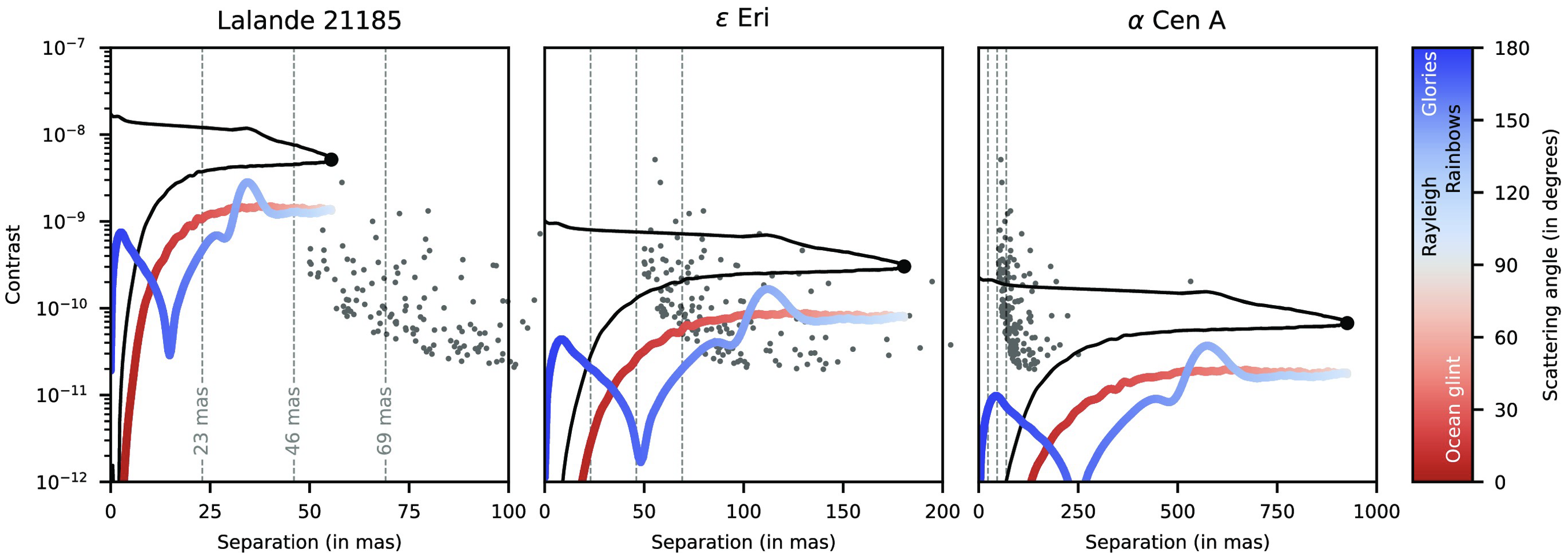}
    \caption{Planet/star contrast (at 670 nm) and separation throughout the orbit of three potential HWO targets, assuming they were Earth twins. In each panel the solid black line shows photometric contrast while the colored line shows the polarimetric contrast. The planetary contrast is always greatest at superior conjunction, when we see the illuminated side of the planet, but this necessarily occurs at small projected separations. The photometric contrast at quadrature is denoted with a black dot; at this point the angular separation is maximal and the contrast is approximately 3$\times$ lower than at superior conjunction. The photometric contrast at crescent phases (the bottom branch of the black curve) is roughly a factor of two lower than at quadrature, while the polarimetric contrast is about a factor of three lower (although the technical challenges of polarimetry are different than those of photometry, so greater contrast does not directly translate to harder measurements). The technical challenge of glint measurements is primarily in the inner working angle, which eliminates certain potential targets (vertical dashed lines denote three notional IWA).  \citep[Figure from][]{Vaughan2023} 
}
    \label{fig:fig2}
\end{figure*}

The most robust way to establish the presence of liquid surface water on an exoplanetary body is via specular reflection, commonly known as \emph{glint} \citep[e.g.,][]{McCullough2006, Williams2008}. The top panel of \autoref{fig:fig1} shows phase-resolved reflectance of a planet with oceans, including the tell-tale shoulder of glint at crescent phases (on the right). The technical challenges with this approach are that it implies measurements of the planet at crescent phases (which are not always accessible, depending on orbital inclination), which requires a small inner working angle (\autoref{fig:fig2}). In the likely event that continuous monitoring spanning months is infeasible, then absolute instrument stability of 10\% or better on timescales of months would be required. Moreover, even if specular reflection increases the planet’s apparent albedo, the flux of the planet will be 3–10 times lower than at gibbous phases (\autoref{fig:fig2}). Clouds are a confounding factor, since forward scattering would also increase the planet’s apparent albedo at crescent phases \citep{Robinson2010}. Moreover, there is a partial degeneracy between the 2D surface map of the planet and its scattering phase function \citep{Cowan2012}. 

The linearly polarized shoulder of specular reflection (bottom panel of \autoref{fig:fig1}) would be less ambiguous, motivating polarimetric capabilities on HWO. Moreover, polarization could reveal many other key atmospheric properties, including identifying water clouds via the distinctive rainbow \citep{GarciaMunoz2015, Vaughan2023, GoodisGordon2025}. 

Detecting ocean glint, however, does not necessarily distinguish between an Earth-like surface and a completely water-covered ocean world. The details of a planet’s ocean coverage are critical, as a completely ocean-covered world might not have a silicate weathering thermostat \citep{Abbot2012, Nakagawa2020, Kite2018, Krissansen-Totton2021}, nor might it be a plausible place for abiogenesis (no warm tidal pools, impossibly deep hydrothermal vents) and the evolution of advanced life (no land-based life, no continental shelves for hydrocarbons to accumulate and enable an industrial revolution, \emph{add your anthropocentric musings here}).  Completely water-covered worlds could also be susceptible to abiotic oxygen accumulation and thus need to be identified or ruled out before oxygen biosignatures can be interpreted \citep{2025arXiv250621790U}. 

\subsection{Ocean Mapping}

Alternatively, surface oceans can be identified at gibbous phases or near quadrature via rotational variability \citep{Ford2001} and mapping \citep{Cowan2009, Cowan2018}. Specifically, oceans would appear as dark regions in the map since light can penetrate into the ocean and only the blue wavelengths scatter back out. Mapping at crescent phases is attractive in theory because of the narrower illuminated and visible sliver of planet enhances the spatial resolution attainable with rotational mapping, but in practice it is harder because a) the planetary flux is lower and b) scattering phase functions of many surfaces, including clouds, become distinctly non-Lambertian.  

\begin{figure*}[ht!]
    \centering
    \includegraphics[width=0.95\textwidth]{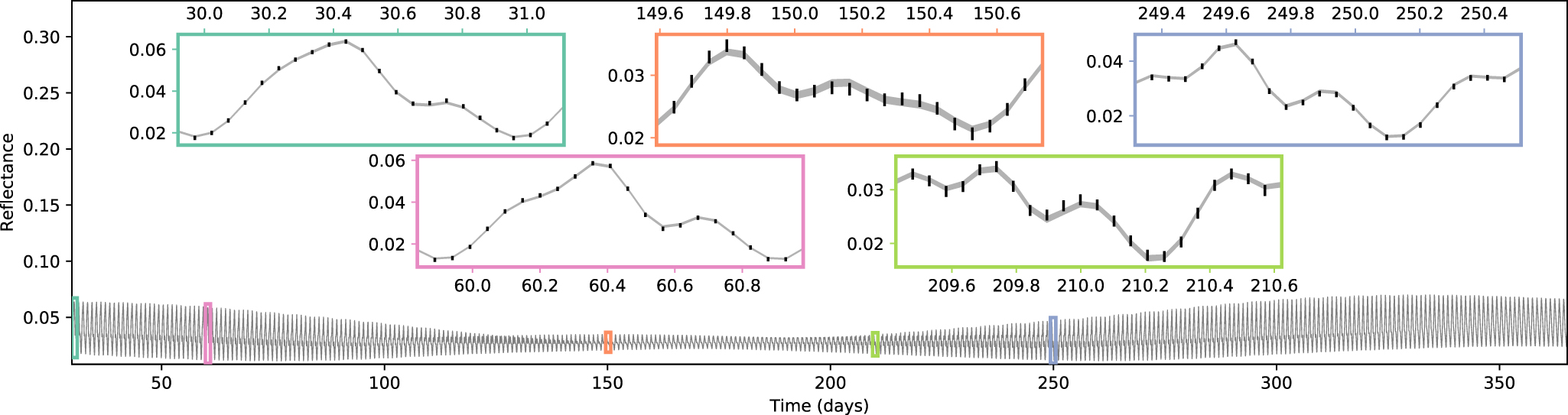}
    \caption{Simulated full-orbit light curve—and resulting fit—for an Earth analog with $0^{\circ}$ orbital inclination (face-on orbit) and 90° rotational obliquity. The simulated data and measurement errors are shown in the five insets, representing five epochs, each lasting one rotation, with a one-hour cadence, collected over the course of a year. The shaded region in each inset shows the central 90\% posterior credible interval for the reconstructed light curve. A fit to the rotational variability at each epoch yields a longitudinal map of the planet, while a joint fit to multiple epochs yields a two-dimensional map of the planet (e.g., \autoref{fig:fig4}). \citep[Figure from][]{Farr2018}}
    \label{fig:fig3}
\end{figure*}

\begin{figure*}[ht!]
    \centering
    \includegraphics[width=0.59\textwidth]{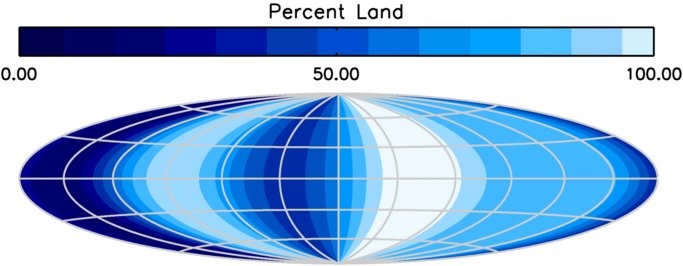} 
    \includegraphics[width=0.39\textwidth]{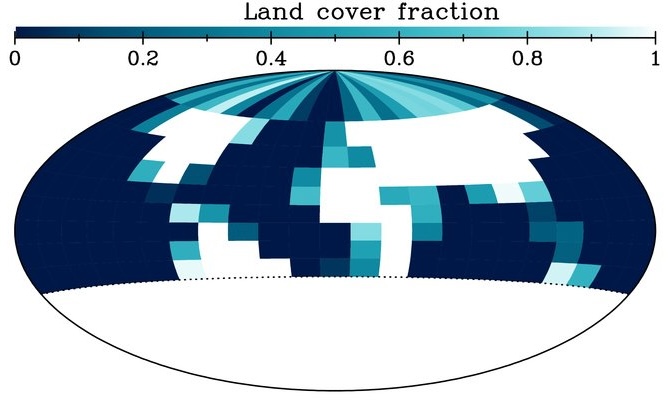}
    \caption{\emph{Left}: A one-dimensional color map of Earth based on 24 hours of disk-integrated multi-band photometry (from \citealp{Cowan2009}). \emph{Right}: A two-dimensional surface map recovered from simulated full-orbit multi-band observations of a cloud-free Earth twin (from \citealp{Kawahara2010}); in this simulation the observer was at northern latitudes and hence could not map the southern portion of the planet. In either of these maps, one can identify the major landforms and oceans of Earth, plausibly identifying this planet as habitable. 
}
    \label{fig:fig4}
\end{figure*}

The order of operations for surface mapping might be:
\setlist{nolistsep}
\begin{enumerate}[noitemsep]
\item Identify photometric rotational variability (\autoref{fig:fig3}). This will require relative instrument stability on timescales of days.
\item Measure the planet’s rotational period \citep{Ford2001, Palle2008, Oakley2009, GarciaMunoz2015, Jiang2018}, with implications for its formation \citep{Miguel2010}, evolution \citep{Barnes2018}, and climate state \citep{Yang2014}.
\item Produce a longitudinal map of the planet (left panel of \autoref{fig:fig4}) by inverting the rotational lightcurve \citep{Cowan2009, Cowan2011}, via time-resolved spectroscopy, multi-band photometry, or single-band photometry (although the latter struggles to distinguish between clouds and continents: \citealp{Oakley2009, Cowan2009, Teinturier2022}).
\item It is also possible to measure the planet’s obliquity \citep{Schwartz2016, Kawahara2016, Farr2018, Nakagawa2020} and produce a two-dimensional map of the planet (latitude and longitude) by inverting rotational lightcurves obtained at multiple epochs throughout the planet’s orbit (right panel of \autoref{fig:fig4}, \citealp{Kawahara2010}). Two dimensional mapping is only feasible if the planet has a significant obliquity so that different latitudes are illuminated at different orbital phases. A confounding factor here is the variability in cloud cover from one epoch to another.
\item Identify the surface features in the 1D or 2D ---ideally multi-wavalength--- map of the planet \citep{2013ApJ...765L..17C,2017AJ....154..189F}. Oceans could be associated with dark regions, especially if spectra of these regions exhibit a flat spectrum. Continents are more reflective and redder. Unfortunately, some rocks are darker than others and could be conflated with liquid water oceans (e.g., the Moon’s mare and highlands). On water-worlds, on the other hand, the dominant contrast might be between open ocean and ice-covered water \citep{Pierrehumbert2011, Hu2014}.  Clouds, meanwhile, are even more reflective than snow-free continents and tend to appear gray at optical wavelengths. Lastly, it might be possible to detect and map the red edge of chlorophyll, a biosignature \citep{Fujii2010}.
\end{enumerate}

Rotational mapping only works for planets with large-scale longitudinal variations in albedo. Plate tectonics seems to naturally produce large continents, but their distribution on the planet is random.  We must therefore expect that some Earth-like planets will not be mappable due to an unfortunate continental configuration (e.g., super-continent at a pole).  Such worlds would still be amenable to glint measurements if their orbit is sufficiently edge-on, however.

Spectral unmixing of surface types is a less exact science than spectral retrieval of exoplanet atmospheres. Liquids and solids do not exhibit the quantum ``barcode'' of electronic and rotational-vibrational energy transitions. Hence, even when we are able to measure the spectra of different regions of an exoplanet, our interpretation of these regions will remain ambiguous.     

A hybrid approach is to map the specular reflection on a planet \citep{Lustig-Yaeger2018}. This approach has the advantage of detecting not only oceans (which glint), but also continents (which don’t). From a technical standpoint, this approach is challenging because it requires high-cadence, high-SNR photometry or polarimetry at crescent phases (requiring a tight inner working angle), when the planetary flux is lower. Patchy clouds are a confounding factor since they can mask underlying water, or forward scatter starlight themselves. The rotational modulation of glint is a double-edged-sword, however: short exposures may miss ocean glint due to clouds or continents rotating through the glint spot and the glint signature in longer exposures could be diluted by non-specular regions on the planet \citep{Schneider2024}.  

\section{Physical Parameters}

\noindent \textbf{\textit{All of the proposed techniques for identifying surface water rely on precise, time-resolved measurements of the planet’s reflectance.}} \\ 

Since it is currently impossible to predict the rotation rate or surface appearance of terrestrial exoplanets, we simply design a survey that would be sensitive to Earth-like planets. The Earth rotates on its axis every 24 hrs and exhibits photometric rotational variability of 15–30\%.  Since the Earth’s albedo is roughly 30\%, the rotational variations of the Earth are of only 10\% in the full apparent albedo scale of zero to unity. In order to detect these variations at 10 sigma, we need 1\% photometry with respect to the reflected light figure of merit, $(R_{\rm p}/a)^2$.  

Rotational mapping requires instrument stability on the timescale of the planet’s rotational period.  This timescale will not be known a priori, but is plausibly within an order of magnitude of Earth’s 24 hours.  Although instrument stability on these shorter timescales should not pose a challenge for HWO, the photometric precision may be a challenge, especially for rapidly-rotating worlds. One would need at least 4 integrations per rotation to measure the planet’s rotational period, let alone map its surface. Hence, for an Earth twin, one would need 1\% photometry in a 6 hour integration at quadrature. If a planet does not have variable clouds, then observations from multiple rotations could be combined; this strategy requires less photometric precision but is risky since ocean-bearing worlds likely have a hydrological cycle and hence time-variable clouds. In practice, most published numerical mapping exercises have assumed 1\% multi-band photometry every hour. If HWO performance is significantly worse than this, then new simulations will have to be performed to quantify the feasibility of surface mapping.   
 
The timescales, however, depend on the approach.  For glint detection, measurements span the orbit of the planet, either in the form of a long, continuous campaign (with implications for the telescope’s field of regard, among other things) or multi-epoch measurements spaced months apart (requiring an instrument with excellent absolute stability). Polarimetric or spectroscopic glint detection, being differential measurements, are better in this respect: it is easier to compare the polarization fraction or color of the planet at two epochs, than to compare its brightness. For either photometric or polarimetric glint detection, the orbit needs to be sufficiently inclined that the planet exhibits significantly different phases, but in principle a mere two epochs could be sufficient to identify a non-Lambertian scattering phase function, and the exposure times can be longer than the planetary rotation period.   

Given the primary HWO goals of obtaining reflectance spectra of directly-imaged temperate terrestrial exoplanets, the main additional challenges for surface water detection via glint are: 
\setlist{nolistsep}
\begin{itemize}[noitemsep]
    \item NEED: Small inner working angle (30 mas; \autoref{fig:fig7}, below)
    \item NEED: Photometric stability of 10\% on months-long timescales OR Polarimetric coronagraphic capabilities (WANT: simultaneous Q and U)
\end{itemize}
While for surface mapping the additional challenges are: 
\begin{itemize}
\itemsep\smallspace
    \item NEED: Photometric precision of 1\% in few hr integration 
    \item WANT: Simultaneous multi-band photometry
\end{itemize} 

In practice, a couple of synergies with other HWO science objectives may arise.  For example, multi-epoch imaging will likely be needed for orbit determination of any potential temperate terrestrial targets \citep{Guimond2019}. Provided that the planetary reflectance can be precisely measured, then it could serve both to refine the planet’s orbit \citep{Bruna2023} and quantify the presence of glint. Moreover, long integration times will likely be required to build up signal-to-noise and wavelength coverage for spectroscopy of putative temperate terrestrial targets. Provided that these spectrographs offer flux stability, these observations could be binned in wavelength to obtain time-resolved photometry to map the surface of the planet.

\begin{table*}[tbh!]
    \centering
    \caption[Performance Goals]{Performance goals for ocean glint.}
    \label{tab:performance_glint}
    \begin{tabular}{lcccc}
        \noalign{\smallskip}
        \hline
        \noalign{\smallskip}
        \thead[l]{Physical \\ Parameter} & {State of the Art} & {Incremental Progress} & {Substantial Progress} & {Major Progress} \\
        \noalign{\smallskip}
        \hline
        \hline
        \noalign{\smallskip}
        \makecell[l]{Contrast Ratio} & $10^{-8}$ (Roman) ELTs  & $10^{-9}$ & $10^{-10}$ & $10^{-11}$ \\ 
        \hline
        \makecell[l]{Planet Phase \\ Angle} & N/A & \makecell{$120^{\circ}$ \\ (Barely crescent)} & \makecell{$150^{\circ}$ \\ (Nominal Crescent)} & \makecell{$160^{\circ}$ \\ (Small Crescent)} \\ 
        \hline
        \makecell[l]{Phase Angle \\ Uncertainty} & {N/A (Roman)} & \makecell{$\pm20$ degrees} & \makecell{$\pm10$ degrees} & \makecell{$\pm5$ degrees} \\ 
        \hline 
        \makecell[l]{Deviation from \\ Lambertian} & (Roman) & \makecell{$\pm0.1$ in $g$} & \makecell{$\pm0.05$ in $g$} & \makecell{$\pm0.01$ in $g$} \\ 
        \hline 
        \makecell[l]{Number of \\ Planets} & N/A & \makecell{1 exo-Earth} & \makecell{10 exo-Earths} & \makecell{20 exo-Earths} \\ 
        \noalign{\smallskip}
        \hline
    \end{tabular}
\end{table*}

\begin{figure}[th!]
    \centering
    \includegraphics[width=0.47\textwidth]{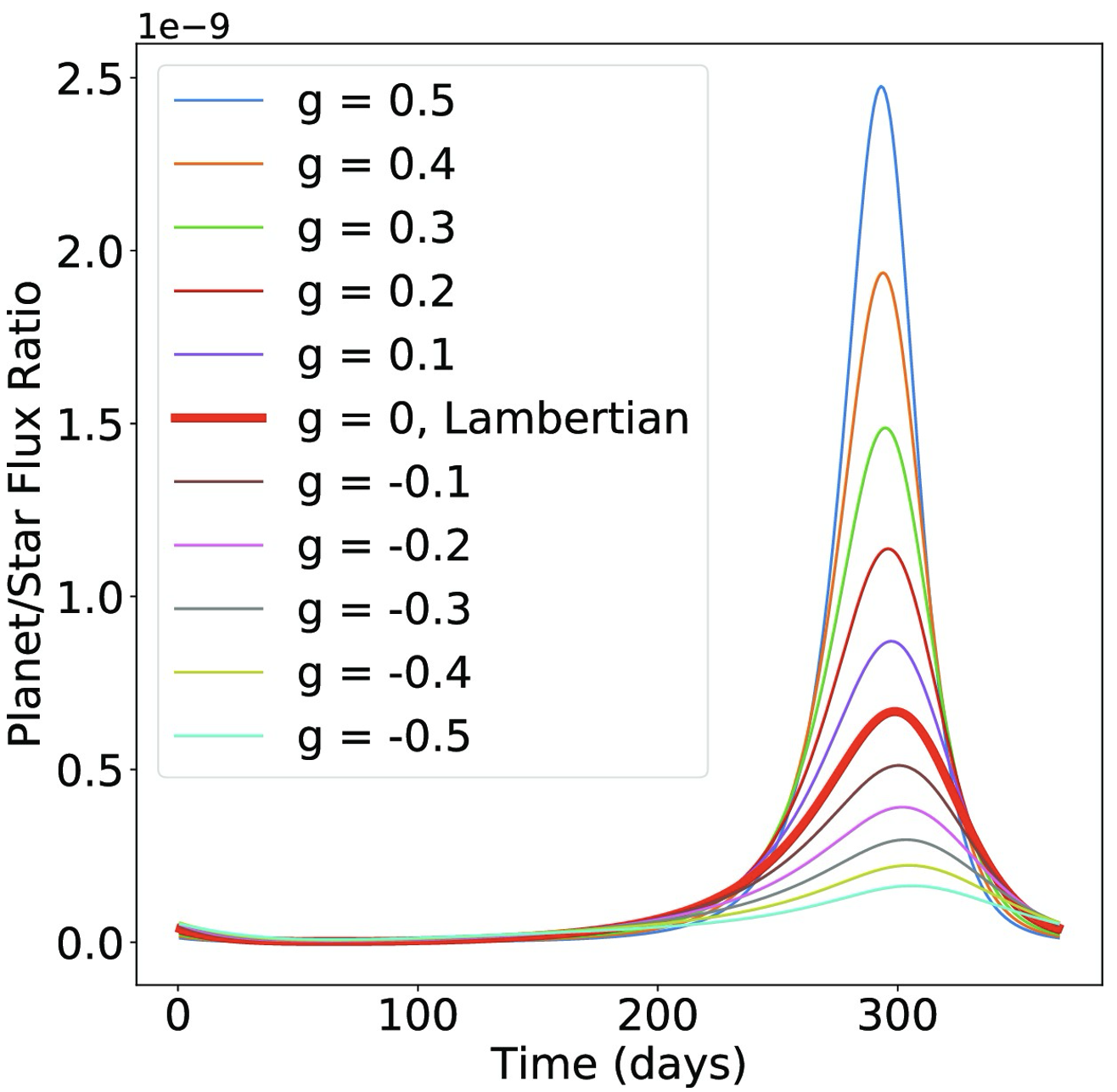}
    \caption{Example reflected light phase curves for an eccentric planet as a function of the Henyey–Greenstein scattering asymmetry factor $g$ (from \citealp{Bruna2023}). }
    \label{fig:fig5}
\end{figure}

\autoref{tab:performance_glint} provides the performance goals for the ocean glint science case. Physical parameters are listed along with the current or near term state of the art capabilities and the goals for incremental, substantial, and major progress. 
Contrast ratio is driven by the crescent phase Earth. Phase angle must be known to calculate the phase dependent albedo. Deviation from Lambertian can be quantified by the precision on scattering asymmetry factor g in the Henyey–Greenstein scattering law (\autoref{fig:fig5}). Alternatively, the phase function of an exoplanet could be modeled as an admixture of diffuse and specular reflection; we wish to know the precision with which the specular fraction can be constrained. In principle, the precision of the phase angle and deviation from Lambertian could be increased by obtaining measurements at more phase angles or increasing the time on target at any phase angle.   

\autoref{tab:performance_mapping} provides the performance goals for the ocean mapping science case. Physical parameters are listed along with the current or near term state of the art capabilities and the goals for incremental, substantial, and major progress. 

\begin{table*}[tbh!]
    \centering
    \caption[Performance Goals]{Performance goals for ocean mapping.}
    \label{tab:performance_mapping}
    \begin{tabular}{lcccc}
        \noalign{\smallskip}
        \hline
        \noalign{\smallskip}
        \thead[l]{Physical \\ Parameter} & {State of the Art} & {Incremental Progress} & {Substantial Progress} & {Major Progress} \\
        \noalign{\smallskip}
        \hline
        \hline
        \noalign{\smallskip}
        \makecell[l]{Rotational \\ Variability} & \makecell{Gas giants with \\ Roman or ELTs} & \makecell{5 sigma variability \\ of a few Super-Earths} & \makecell{5 sigma variability \\ of a few Earths} & \makecell{8 sigma variability \\ of a dozen Earths} \\ 
        \hline 
        \makecell[l]{Rotational \\ Period} & \makecell{A few giants\\ from Roman} & \makecell{Rotation periods to 10\% \\ for a few Super-Earths} & \makecell{Rotation periods to 10\% \\ for a few Earths} & \makecell{Rotation periods to 10\% \\ for a dozen Earths} \\ 
        \hline
        \makecell[l]{Albedo \\ Inhomogeneity} & \makecell{A few giants \\ from Roman} & \makecell{5 sigma inference of \\ multiple surface \\ components on SEs} & \makecell{5 sigma inference of \\ multiple surface \\ components on Earths} & \makecell{8 sigma inference of \\ multiple surface \\ components on Earths} \\
        \hline
        \makecell[l]{Surface \\ Colors} & N/A & \makecell{Broadband colors of \\ surfaces to 20\% \\ on Super-Earths} & \makecell{Broadband colors of \\ surfaces to 20\% \\ on Earths} & \makecell{Broadband colors of \\ surfaces to 10\% \\ on Earths} \\ 
        \noalign{\smallskip}
        \hline
    \end{tabular}
\end{table*}

\section{Description of Observations}

\begin{figure}[h!]
    \centering
    \includegraphics[width=0.47\textwidth]{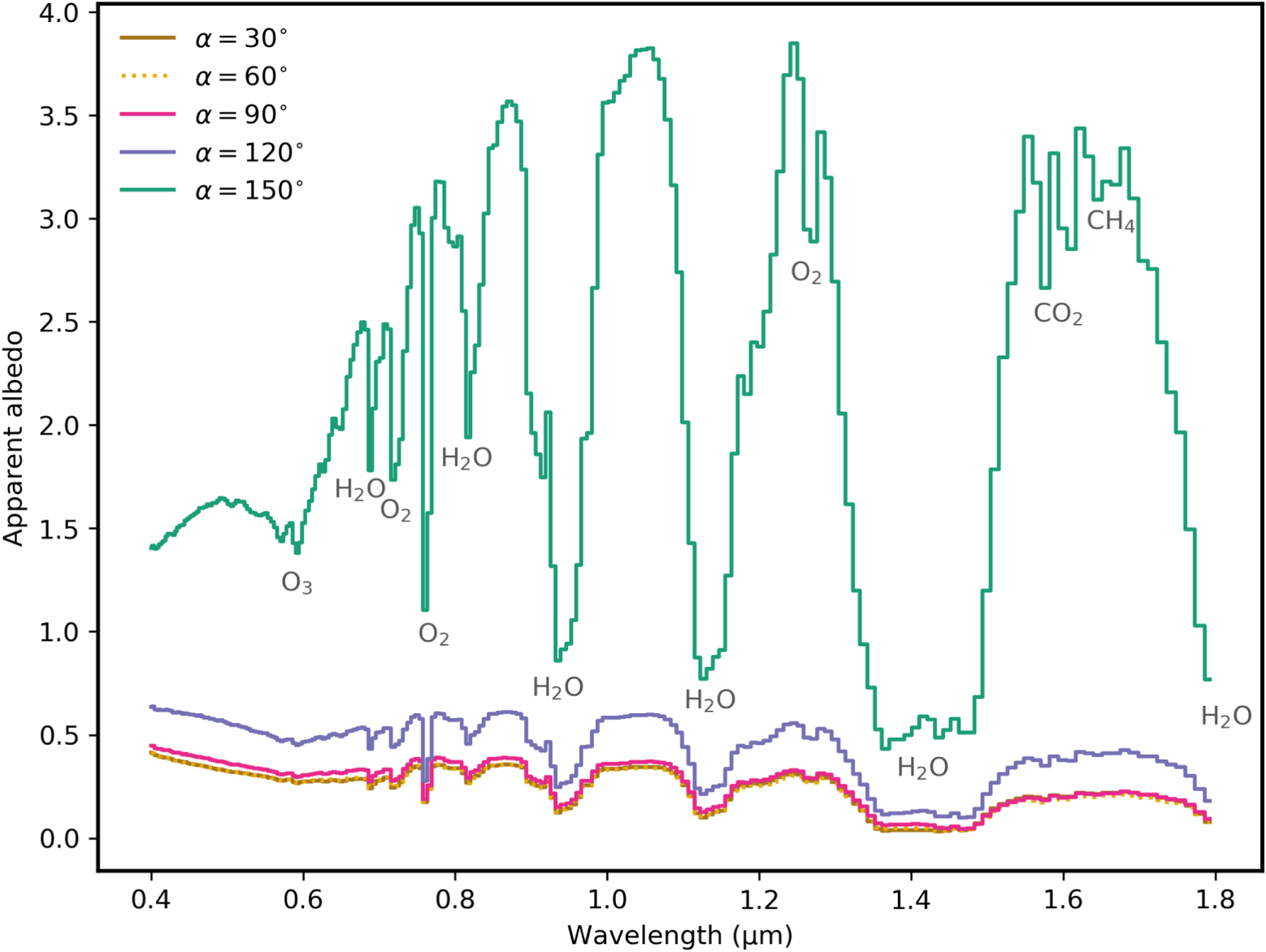}
    \caption{Example apparent albedo spectra from the VPL Earth model across a range of phase angles and degraded to a resolving power of 140 (from \citealp{Ryan2022}). Observations between water bands have the strongest sensitivity to the planet surface. At crescent phase angles ${>}120^{\circ}$ Earth’s apparent albedo increases (as a deviation from Lambertian scattering) due primarily to ocean glint. }
    \label{fig:fig6}
\end{figure}

\begin{figure*}[ht!]
    \centering
    \includegraphics[width=0.95\textwidth]{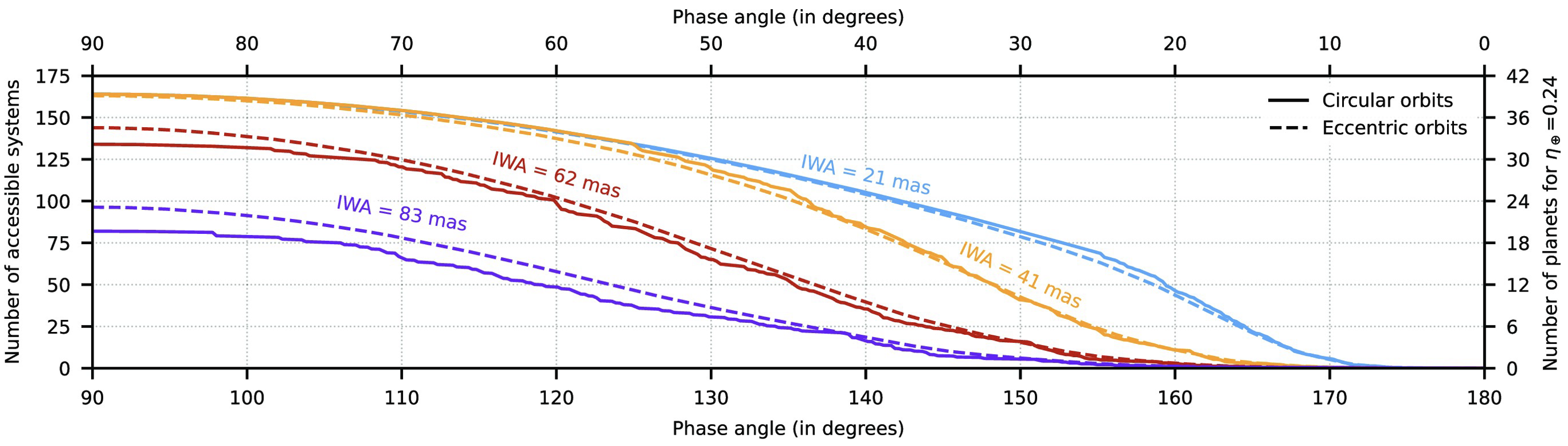}
    \caption{Number of accessible exo-Earths at crescent phases for different inner working angles (IWA). From \citet{Vaughan2023}. }
    \label{fig:fig7}
\end{figure*}

Detection of surface water involves high-precision time-resolved photometry of the planet in coronagraphic mode to block out its host star. These measurements could also be made using spectroscopy if the flux calibration is sufficiently stable. In any case, these observations should be performed at red optical wavelengths, where Rayleigh scattering from the atmosphere is minimal and the albedo contrast between land and water is greatest for mapping. Furthermore, we expect a planet with exposed surface water to have water vapor in its atmosphere, so the sensitivity to surface scattering properties (both glint and mapping) should be made in the spectrum continuum between water bands (e.g., 0.8--1.1\,µm; specifically, 0.8--0.9\,µm and 1.0--1.1\,µm), see \autoref{fig:fig6}. 

\textbf{Glint:} measure the reflectance of the planet at four different orbital phases going from gibbous to crescent (30 degree scattering angle). Each epoch can involve long integration times: a day or a week, if needed.  The key is to obtain high precision reflectance observations of the planet that can be compared between epochs to detect deviations from Lambertian behavior. The four or so epochs can be spaced out by months or years. For maximum impact these observations should be obtained in polarization and with multi-band photometry.

\textbf{Mapping:} measure the reflectance of the planet at high cadence to resolve its rotational variability.  Integrations should be approximately an hour (or shorter for fast rotating planets). Such a campaign must span at least one rotation of the planet; this period will not be known a priori but will likely be hours to days long. Ideally the observations will consist of simultaneous, multi-band photometry (or spectroscopy with a wide wavelength range and high stability).

\autoref{tab:obsreq_glint} and \autoref{tab:obsreq_mapping} provide the observational requirements for the ocean glint and ocean mapping science cases, respectively. Observational parameters are listed along with the current or near term state of the art capabilities and the goals for incremental, substantial, and major progress.

\begin{table*}[ht!]
    \centering
     \caption[Observation Requirements]{Observation requirements for ocean glint science case.}
    \label{tab:obsreq_glint}
    \begin{tabular}{lcccc}
        \noalign{\smallskip}
        \hline
        \hline
        \noalign{\smallskip}
        {Observation} & {State of the Art} & {Incremental Progress} & {Substantial Progress} & {Major Progress} \\
        \noalign{\smallskip}
        \hline
        \noalign{\smallskip}
        \makecell[l]{Wavelength \\ Range} & N/A & 0.4 – 0.9 µm & 0.4 – 1.0 µm & \makecell{0.4 – 1.1 µm \\ (w/ polarimetry)} \\ 
        \hline
        \makecell[l]{Contrast \\ Ratio} & $10^{-8}$ (Roman) & $10^{-9}$ & $10^{-10}$ & $10^{-11}$ \\ 
        \hline
        \makecell[l]{IWA $@$ $\lambda_{\rm max}$ \\ (\autoref{fig:fig7})} & \makecell{150 mas \\ (Roman)} & \makecell{80 mas: HZ  \\ at 120 degrees \\ orbital phase} & \makecell{50 mas: HZ \\ at 150 degrees \\ orbital phase} & \makecell{20 mas: HZ \\ at 160 degrees \\ orbital phase} \\ 
        \hline
        \makecell[l]{Observation \\ Epochs} & Roman? & \makecell{2 spanning \\ 30 degrees} & \makecell{6 spanning \\ 40 degrees} & \makecell{10 spanning \\ 50 degrees} \\
        \hline
        \makecell[l]{SNR} & \makecell{N/A without HWO} & \makecell{10 on spectrum \\ continuum} & \makecell{15 on spectrum \\ continuum} & \makecell{20 on spectrum \\ continuum} \\ 
        \noalign{\smallskip}
        \hline
    \end{tabular}
\end{table*}

\begin{table*}[ht!]
    \centering
     \caption[Observation Requirements]{Observation requirements for ocean mapping science case.}
    \label{tab:obsreq_mapping}
    \begin{tabular}{lcccc}
        \noalign{\smallskip}
        \hline
        \noalign{\smallskip}
        {Observation} & {State of the Art} & {Incremental Progress} & {Substantial Progress} & {Major Progress} \\
        \noalign{\smallskip}
        \hline
        \hline
        \noalign{\smallskip}
        \makecell[l]{Wavelength \\ Range} & N/A & 0.7 - 0.9 µm & 0.5 - 1.1 µm & 0.4 - 1.1 µm \\ 
        \hline
        \makecell[l]{Contrast \\ Ratio} & $10^{-8}$ (Roman) & $10^{-9}$ & $10^{-10}$ & $10^{-11}$ \\ 
        \hline
        \makecell[l]{Uncertainty} & N/A & $0.1 (R_p/a)^2$ & $0.03 (R_p/a)^2$ & $0.01 (R_p/a)^2$ \\ 
        \hline
        \makecell[l]{Observation \\ Baseline} & Roman? & 24 hours & 72 hours & 1 week \\ 
        \hline
        \makecell[l]{Observation \\ Cadence} & Roman? & 6 hrs & 3 hrs & 1 hr \\ 
        \noalign{\smallskip}
        \hline
    \end{tabular}
\end{table*}

\section{Sources of Uncertainties and Future Work Needed}

Glint detection relies on knowing the orbital phase of the planet at each epoch. In practice, the planetary orbit will not be known perfectly. Nobody has yet explored how uncertain planetary orbital parameters would impact inferences of glint. A possible starting point would be to extend the simulations of \citet{Bruna2023} to a) quantify the precision one can obtain on the scattering phase function when the orbital parameters are uncertain, and b) accounting for ancillary constraints on the planetary orbit from stellar radial velocity, astrometry, etc. 

Glint detection would benefit from a quantitative framework for retrieving the scattering phase curve of a planet \citep[possibly a variant of][]{2021NatAs...5.1001H}.  This could be in terms of the scattering parameter g in the Henyey–Greenstein phase function. Alternatively, reflectance could be parameterized as a sum of diffuse and specular reflection, e.g., at orbital phase such-and-such, $30\pm 5$ percent of the visible and illuminated region of the planet is covered in specularly-reflecting surface, presumably water. 

We have described the glint and rotational mapping cases as distinct, but in practice we would have to account for rotational variability when retrieving the scattering phase function.  This could be particularly challenging for planets that are slowly rotating, because even a long integration of 24 hrs might be sampling only one part of the planet and hence might be biased. Simulations building on \citet{Lustig-Yaeger2018} could be used to see how the \emph{a priori} unknown rotation and map of a planet would impact inferences of ocean glint. For full value, such simulations could account for imprecise knowledge of the planetary orbit ---including eccentricity--- and hence orbital phase. 

Rotational mapping simulations have always assumed simultaneous multi-band photometry, which would not be feasible for all coronagraph designs. The impact of contemporaneous (but non-simultaneous) observations should be quantified. For example, would it be acceptable to monitor the planet for one rotation in the blue, followed by one rotation in the red? 

Rotational mapping simulations have assumed that the rotational period of the planet is known a priori due to an initial campaign of photometric monitoring.  In practice, that initial monitoring may be the same data used to establish the orbit of the planet, or to obtain a NIR spectrum showing water absorption. The trade-offs of such data sharing should be explored. At the very least, the mapping retrieval exercise should be tested in scenarios where the rotational period and planetary orbit are not yet known, but rather are being simultaneously fitted.   

There is relatively limited polarization modeling work to draw upon. For example, by building on the work of \citet{Vaughan2023}, assessing the detectability of the scattering features with HWO would help robustly establish such polarization science cases. 

\section{The Need for HWO} The European Extremely Large Telescope will have first light in the late 2020s and at least one US-led ELT is expected in the 2030s. These telescopes, using high resolution near-infrared spectroscopy behind a high contrast imaging system should be able to search for biosignatures on a handful of rocky planets orbiting nearby M-dwarfs.  ELTs may even be able to detect a handful of Earth twins in the thermal infrared. 

The only other telescope being seriously discussed that could characterize large numbers of temperate terrestrial planets orbiting Sun-like (FGK) stars is the Large Interferometer For Exoplanets (LIFE) mission concept \citep{Quanz2022}, and it could only identify surface oceans indirectly. Long-term monitoring of a planet with LIFE might constrain the tempering presence of surface oceans on seasonal temperature variations \citep{Cowan2012}; such climatological constraints would be complementary to the ground truth provided by HWO. Indeed HWO is uniquely capable of identifying surface liquid oceans via their optical properties. Given that discovering an ocean on an exoplanet would confirm its status as a habitable world, this science case is literally the \emph{raison d’être} of the Habitable Worlds Observatory.

\vspace{0.2cm}
{\bf Acknowledgements.} N.B.C. acknowledges support from a Canada Research Chair, NSERC Discovery Grant, and McDonald Fellowship. He also thanks the Trottier Space Institute and l’Institut de recherche sur les exoplanètes for their financial support and dynamic intellectual environment. J.L.Y. acknowledges internal support from Johns Hopkins APL. 


\bibliography{main} 

\end{document}